\newcommand{\diracslash}[1]{#1\llap{/\kern2pt}}

\newcommand{\be}{\begin{equation}}
\newcommand{\ee}{\end{equation}}
\newcommand{\bea}{\begin{eqnarray}}
\newcommand{\eea}{\end{eqnarray}}
\newcommand{\ba}[1]{\begin{array}{#1}}
\newcommand{\ea}{\end{array}}

\documentclass[twocolumn,superscriptaddress,secnumarabic,amssymb,nobibnotes,nofootinbib,aps,pre,showpacs]{revtex4}
\usepackage{graphicx}
\begin{document}

\title{Multiresolution analysis of fluctuations in non-stationary time series through discrete wavelets}
\author{P. Manimaran}
\affiliation{School of Physics, University of Hyderabad,
 Hyderabad 500 046, India}
\author{Prasanta K. Panigrahi}
\affiliation{Indian Institute of Science Education and Research
(Kolkata), Salt Lake, Kolkata 700 106, India} \affiliation{ Physical
Research Laboratory, Navrangpura, Ahmedabad 380 009, India}
\author{Jitendra C. Parikh}
\affiliation{ Physical Research Laboratory, Navrangpura, Ahmedabad
380 009, India}

\date{\today}

\def\be{\begin{equation}}
\def\ee{\end{equation}}
\def\bearr{\begin{eqnarray}}
\def\eearr{\end{eqnarray}}
\def\zbf#1{{\bf {#1}}}
\def\bfm#1{\mbox{\boldmath $#1$}}
\def\hf{\frac{1}{2}}

\begin{abstract}

We illustrate the efficacy of a discrete wavelet based approach to
characterize fluctuations in non-stationary time series. The present
approach complements the multi-fractal detrended fluctuation
analysis (MF-DFA) method and is quite accurate for small size data
sets. As compared to polynomial fits in the MF-DFA, a single
Daubechies wavelet is used here for de-trending purposes. The
natural, built-in variable window size in wavelet transforms makes
this procedure well suited for non-stationary data. We illustrate
the working of this method through the analysis of binomial
multi-fractal model. For this model, our results compare well with
those calculated analytically and obtained numerically through
MF-DFA. To show the efficacy of this approach for finite data sets,
we also do the above comparison for Gaussian white noise time series
of different size. In addition, we analyze time series of three
experimental data sets of tokamak plasma and also spin density
fluctuations in 2D Ising model.

\end{abstract}

\pacs{05.45.Df, 05.45.Tp, 89.65.Gh}

\maketitle
\section{Introduction}
For non-stationary time series, it is of vital importance to
'correctly' separate fluctuations from average behavior (trend), for
studying various properties of real systems having complex dynamics
-- i.e., having many different spatio-temporal scales. Several
techniques have been developed to carry out this separation. Among
these are, de-trended fluctuation analysis and its variants
\cite{peng,net} and the wavelet transform \cite{daub,mall} based
multi-resolution analysis \cite{mani,ran}. These methods and earlier
methods \cite{mandel,arn1} have found wide application in analysis
of correlations and characterization of scaling behavior of
time-series data in, physiology, finance, and natural sciences
\cite{khu,gopi,ple,chen,matia,krs,phand,xu,nico,gu,sade}. Recently,
the relative merits of MF-DFA and a variety of other approaches to
characterize fluctuations have been carried out \cite{jaro}. It is
worth emphasizing that fluctuation analysis and characterization
have been earlier attempted using Haar wavelets, in the context of
bio-medical applications, without the study of scaling behavior
\cite{pkp1,pkp2}.

In this paper, we present a refined version of our earlier procedure
\cite{mani}, where a local averaging procedure is adopted to
accurately separate fluctuations, from the trend, akin to the MF-DFA
approach. It is important to further emphasize that, in our earlier
method, global averaging had been carried out for finding the
fluctuations. This necessitated the use of two wavelet basis to
separate large and small fluctuations. In comparison, here we find
that a single Daubechies wavelet enables one to study small and
large fluctuations together. The effect of correlation of
non-stationary data on the fluctuations are clearly isolated.

Details of the present approach are described in Sec. II. Sec. III
contains analysis of fluctuations of the binomial multi-fractal
model and comparison of results using different approaches. To check
the efficacy of the present method, when the data size is small, we
have carried out a systematic investigation of the scaling exponents
for the Gaussian white noise of different sizes using a number of
wavelets belonging to Daubechies family, which is then compared with
MF-DFA with polynomial fits of different degrees. It is found that
for a small length data the present wavelet based method is better
suited to estimate the scaling exponents. The results illustrate the
correctness of our approach, aside from being theoretically sound
and natural. Subsequently, we carry out analysis of fluctuation data
observed in tokamak plasma depicted in Fig. 1. Results describing
scaling properties of tokamak plasma are given in Sec. IV. We also
analyze the fluctuation characteristics of the spin density of the
2D Ising model. Finally, in Sec. V we summarize our findings and
give some concluding remarks.

\section{Details of modified wavelet approach}

The present wavelet based procedure is explained through the
following steps. Note that the steps are very similar to those in
MF-DFA \cite{net}, except that in order to detrend, we use wavelets
and MF-DFA uses local polynomial fits.

Let $x_t$ (t=1,2,3,...,N) be the time series of length N. First
determine the "profile" (say $Y(i)$), which is cumulative sum of
series after subtracting the mean.
\begin{equation}
Y(i) = \sum_{t=1}^i [x_t - \langle x \rangle], ~~~ i=1,....,N.
\end{equation}

Next, we carry out wavelet transform on the profile $Y(i)$ to
separate the fluctuations from the trend. For this purpose, discrete
wavelets belonging to Daubechies (Db) family is used. It is worth
repeating that these wavelets satisfy the vanishing moment
conditions: $ \int t^m \psi_{j,k}(t) dt = 0 $, where $0 \leq m < n
$. Because of this, the low-pass coefficients keep track of the
polynomial trends in the data. For example, the low-pass
coefficients in Db-4, Db-6 and Db-8, retain polynomial trend which
are linear, quadratic and cubic respectively. Hence, reconstruction
using these low-pass coefficients alone is quite accurate in
extracting the local trend, in a desired window size. The
fluctuations are then extracted at each level by subtracting the
obtained time series from the original data. Though the Daubechies
wavelets extract the fluctuations nicely, its asymmetric nature and
wrap around problem affects the precision of the values. This is
corrected by applying wavelet transform to the reverse profile, to
extract a new set of fluctuations. These fluctuations are then
reversed and averaged over the earlier obtained fluctuations. These
are the fluctuations (at a particular level), which we consider for
analysis.

The extracted fluctuations are subdivided into non-overlapping
segments $M_s = int(N/s)$ where $s=2^{(L-1)}W$ is the wavelet window
size at a particular level (L) for the chosen wavelet. Here $W$ is
the number of filter coefficients of the discrete wavelet transform
basis under consideration. For example, with Db-4 wavelet with 4
filter coefficients, $s=4$ at level 1 and $s=8$ at level 2 and so
on. It is obvious that some data points would have to be discarded,
in case $N/s$ is not an integer. This causes statistical errors in
calculating the local variance. In such cases, we have to repeat the
above procedure starting from the end and going to the beginning to
calculate the local variance.

The $q^{th}$ order fluctuation function, $F_q(s)$ is obtained by
squaring and averaging fluctuations over all segments:
\begin{equation}
F_q(s) \equiv  \{ \frac {1}{2 M_s} \sum_{b=1}^{2 M_s} [
F^2(b,s)]^{q/2}  \}^{1/q}.
\end{equation}

Here 'q' is the order of moment that can take any real value. The
above procedure is repeated for variable window sizes for different
value of q (except q=0). The scaling behavior is obtained by
analyzing the fluctuation function,
\begin{equation}
F_q(s) \sim s^{h(q)},
\end{equation}
in a logarithmic scale for each value of q. If the order $q = 0$,
direct evaluation of Eq. (2) leads to divergence of the scaling
exponent. In that case, logarithmic averaging has to be employed to
find the fluctuation function;

\begin{equation}
F_q(s) \equiv exp \{ \frac {1}{2 M_s} \sum_{b=1}^{2 M_s} ln [
F^2(b,s)]^{q/2} \}^{1/q}.
\end{equation}

As is well-known, if the time series is mono-fractal, the h(q)
values are independent of q. For multifractal time series, h(q)
values depend on q. The correlation behavior is characterized from
the Hurst exponent ($H=h(q=2)$), which varies from $0 < H < 1$.
For long range correlation, $H > 0.5$, $H=0.5$ for uncorrelated
and $H <0.5$ for long range anti-correlated time series.

The behavior of fluctuations extracted through multifractal
detrended fluctuation analysis  and fluctuations obtained using
wavelet transform are shown in Fig. 2 and Fig.3 respectively. We
note that the fluctuations extracted from the two methods differ at
the boundaries.

\section{Analysis of Binomial multifractal model}

For the multifractal time series generated through the binomial
multifractal model \cite{feder,barba,pei}, a series of $N =
2^{n_{max}}$ numbers $x_j$, with $j=1,...,N$, is defined by

\be x_j = a^{n(j-1)}(1-a)^{n_{max}-n(j-1)},\ee

where $0.5 < a < 1$ is a parameter and $n(j)$ is the number of
digits equal to $0$ or $1$ in the binary representation of the
index $j$. The scaling exponent $h(q) = \frac{1}{q} - \frac{ln
[a^q + (1-a)^q]}{q ln(2)}$ and $ \tau(q) = \frac{-ln [a^q +
(1-a)^q]}{ln(2)}$ can be calculated exactly in this model. These
would be compared with numerical results obtained through wavelet
analysis, for illustrating the efficacy of our procedure.

In our wavelet based analysis, profile of the binomial
multifractal model time series has been subjected to a multi-level
wavelet decomposition. The length of the data should be $2^N$,
otherwise constant padding is added at the ends.

In Table-1, we give the $h(q)$ values for various $q$, obtained
from analytical results, MF-DFA and wavelet (Db-8) based method
for binomial multifractal series. In MF-DFA, we have used a
quadratic polynomial fit for extracting fluctuations. It is clear
from the given tables, that the wavelet estimate of the $h(q)$
exponent for the binomial multifractal time series, is extremely
reliable.

\begin{table}
\centering
\begin{tabular}{|c|c|c|c|c|}
\hline \hline
  q & $h(q)_{BMFS_a}$ & $h(q)_{BMFS_s}$ & $h(q)_{BMFS_w}$ \\
  \hline \hline

  -10    &1.9000    &1.9304    &1.8991\\
   -9    &1.8889    &1.9184    &1.8879\\
   -8    &1.8750    &1.9032    &1.8740\\
   -7    &1.8572    &1.8837    &1.8560\\
   -6    &1.8337    &1.8576    &1.8319\\
   -5    &1.8012    &1.8210    &1.7981\\
   -4    &1.7544    &1.7663    &1.7473\\
   -3    &1.6842    &1.6783    &1.6641\\
   -2    &1.5760    &1.5397    &1.5218\\
   -1    &1.4150    &1.3939    &1.3828\\
    0    &     0    &1.2030    &1.2163\\
    1    &1.0000    &0.9934    &1.0091\\
    2    &0.8390    &0.8312    &0.8453\\
    3    &0.7309    &0.7234    &0.7359\\
    4    &0.6606    &0.6538    &0.6649\\
    5    &0.6139    &0.6075    &0.6177\\
    6    &0.5814    &0.5753    &0.5848\\
    7    &0.5578    &0.5519    &0.5610\\
    8    &0.5400    &0.5343    &0.5430\\
    9    &0.5261    &0.5205    &0.5290\\
   10    &0.5150    &0.5095    &0.5178\\
  \hline \hline
\end{tabular}
\caption{The $h(q)$ values of binomial multi-fractal series (BMFS)
computed analytically ($BMFS_a$), through MF-DFA ($BMFS_s$) and
wavelet ($BMFS_w$) approach, Db-8 wavelet has been used.}
\end{table}

In our earlier method, we had used two different wavelets for
analyzing  the large and small fluctuations. We see that, for the
computer generated BMF model time series, the present method, with
only a single wavelet, is quite efficient for characterizing the
correlation properties and multifractal behavior of non-stationary
time series. Our results compare very well with the analytical
results and MF-DFA calculation. It is worth pointing out that, we
have carried out analysis using Daubechies wavelets of various
order. It was found that Db-8 performs the best. The improvement
with higher wavelet is minimal.

For the purpose of finding out the efficacy of the present method
when the data size is small, we now study the Gaussian white noise
of lengths 1000, 5000, 10000 and 50000 points. These are analyzed
through Daubechies 4 (Db-4) to Daubechies 8 wavelets. For the
purpose of MF-DFA, we have computed the trend using linear,
quadratic and cubic polynomial fits. The results are depicted in
Table II, III, IV and V. It is observed that for smaller size data,
wavelet based method is quite effective in estimating the scaling
exponent. In this approach, one wavelet basis is found to be
effective in capturing both smaller and larger fluctuations as
compared to the earlier approach. The local procedure adopted here
is responsible for this improvement.

\section{Analysis of Experimental and synthetic data sets}
We have analyzed three sets of experimentally observed time series
of variables in ohmically heated edge plasma in Aditya tokamak
\cite{jos}. The time series are i) ion saturation current, ii) ion
saturation current when the probe is in the limiter shadow, and
iii) floating potential, 6mm inside the main plasma. Each time
series has about 24,000 data points sampled at 1MHZ \cite{jha}.
These are shown in Fig. 1. The study of fluctuations play an
important role in our understanding of turbulent transport of
particles and heat in the plasma.

\begin{figure}
\centering
\includegraphics[width=3in]{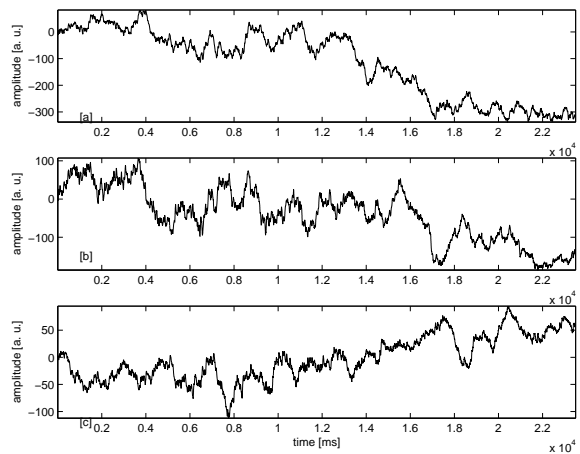}
\caption{Time series of (a) ion saturation current, (b) floating
potential, 6mm inside the main plasma and (c) ion saturation
current, when the probe is in the limiter shadow. Each time series
is of approx. 24,000 data points.}
\end{figure}

\begin{figure}
\centering
\includegraphics[width=3in]{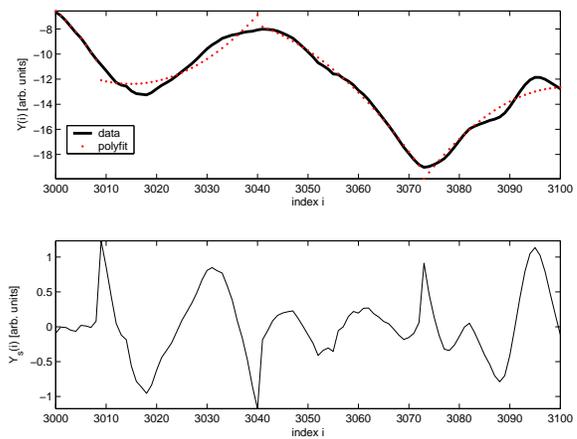}
\caption{Fluctuations extracted from the time series of ion
saturation current, when the probe is in the limiter shadow using
MF-DFA (window size is 32).}
\end{figure}

\begin{figure}
\centering
\includegraphics[width=3in]{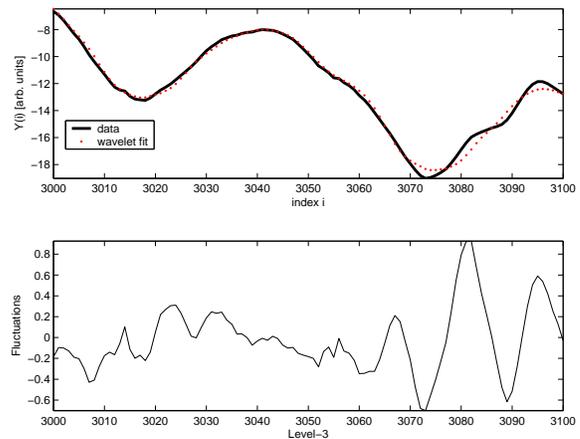}
\caption{Fluctuations extracted from the time series of ion
saturation current, when the probe is in the limiter shadow, at
level-3 using Db-8 wavelet (window size 32).}
\end{figure}

Fluctuation function $F_q(s)$ for various values of q, for the
time series of three experimental data sets are computed using
Db-8 wavelet. In Fig. 4, we have shown $F_q(s)$ versus $s$ of the
time series of ion saturation current, when the probe is in the
limiter shadow.

We have calculated scaling exponents $h(q)$ and $\tau(q)=qh(q)$ for
various $q$ values. All the three time series exhibit non-linear
(fractal) behavior for $h(q)$ values, as a function of $q$. From the
measured Hurst exponent, it was found that the three time series
possess long range correlations, whereas for the shuffled time
series, the same is absent. However the shuffled time series still
shows multifractality, which is clearly seen in Table-VI. Here, the
$h(q)$ values of time series and shuffled time series of i) ion
saturation current (IS), ii) ion saturation current, when the probe
is in the limiter shadow (ISC) and iii) floating potential, 6mm
inside the main plasma (FP) are given, with the subscript 's'
referring to shuffled time series.

The multifractal behavior of the experimental data set can also be
studied from the $f(\alpha)$ spectrum. The $f(\alpha)$ values are
obtained from the Legendre transform of $\tau(q)$. Explicitly, $
f(\alpha) \equiv q \alpha - \tau(q)$, where $\alpha \equiv
\frac{d\tau(q)}{dq}$. For monofractal time series,
$\alpha=const.$, whereas for multifractal time series there will
be a distribution of $\alpha$ values. Fig. 5 shows the $f(\alpha)$
spectrum. In the unshuffled data, one observes a broader spectrum,
whereas for the shuffled data, where the correlation is lost, the
same is narrower.

We now proceed to study the scaling behavior of the spin densities
below and at critical temperature for the 2D Ising model, which are
shown in Fig.6, as a function of time. The spin densities are
computed following the procedure described earlier in
Ref.\cite{mani,hwa}. Below the critical temperature, the
fluctuations show Gaussian white noise character. At critical
temperature the spin densities show a multifractal behavior with
long range correlations, as expected from physical ground. Both
these aspects are clearly shown in Fig. 7.

\begin{figure}
\includegraphics[width=3.2in]{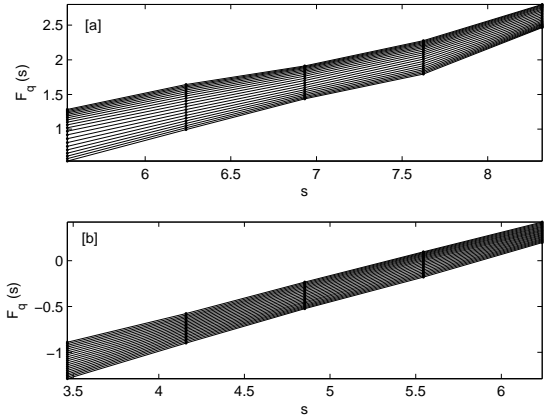}
\caption{The log-log plot of fluctuation function $F_q(s)$ vs $s$
for various values of $q$ (a) for the time series of ion
saturation current, when the probe is in the limiter shadow and
(b) for its shuffled time series, using Db-8 wavelet.}
\end{figure}

\begin{figure}
\includegraphics[width=2.7in]{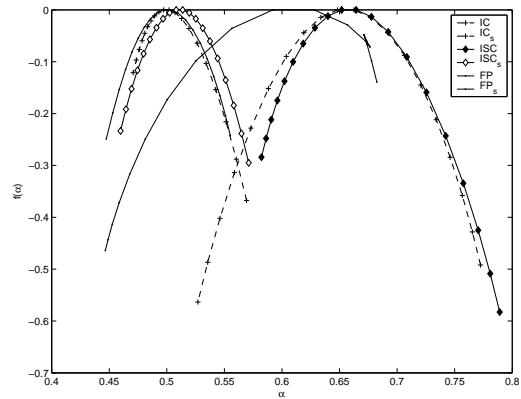}
\caption{The multi-fractal behavior of different type of
experimental data sets is shown through the $f(\alpha)$ spectra
for unshuffled and shuffled time series. Here the subscript 's' in
IC, ISC and FP refers to the shuffled time series. One observes a
broader spectra for the correlated, unshuffled case.}
\end{figure}

\begin{figure}
\includegraphics[width=2.7in]{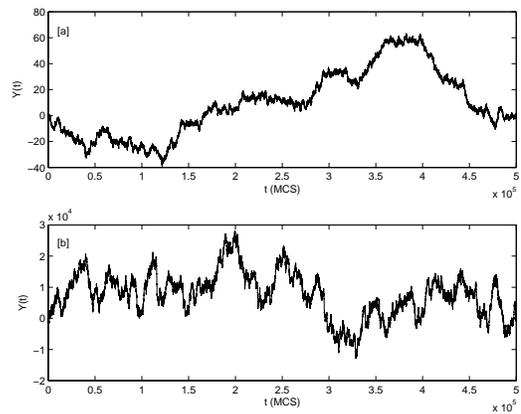}
\caption{The integrated time series of average spin densities after
subtraction of the mean, [a] at $T = 1.0$, below $T_c$ and [b] $T_c
= 2.27$. The difference in behavior of the fluctuations at different
temperatures is clearly visible.}
\end{figure}

\begin{figure}
\includegraphics[width=2.7in]{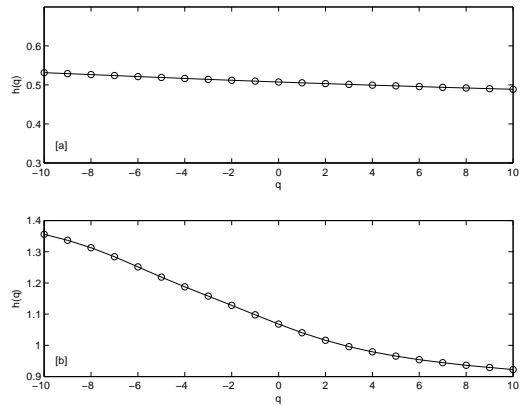}
\caption{For time series [a] Below $T_c$, $h(q)$ shows linear
behavior for different values $q$, indicating a monofractal behavior
and [b] the non-linear behavior of $h(q)$ for different values of
$q$ at $T_c$, shows clearly the long-range correlation and
multifractal nature.}
\end{figure}

\begin{table}
\centering
\begin{tabular}{|c|c|c|c|c|c|c|}
  \hline\hline
  q & Db4 & Db6 & Db8 & Linear & Quadratic & Cubic\\
  \hline\hline
 -10&0.59008&0.58762& 0.6018&0.67735& 0.5487&0.57551\\
 -9&0.58467&0.58343&0.59477& 0.6665&0.54552&0.57085\\
 -8&0.57877&0.57889& 0.5876&0.65406&0.54215&0.56572\\
 -7&0.57246&0.57401& 0.5805&0.63991& 0.5386&0.56003\\
 -6&0.56591& 0.5689&0.57376&0.62404&0.53487&0.55371\\
 -5&0.55936&0.56368&0.56763& 0.6066&0.53096& 0.5467\\
 -4&0.55318&0.55855&0.56228&0.58806&0.52689&0.53896\\
 -3& 0.5477& 0.5537& 0.5578&0.56905&0.52271&0.53052\\
 -2&0.54321&0.54924&0.55419&0.55027&0.51847& 0.5215\\
 -1&0.53981&0.54508&0.55133&0.53218&0.51426&0.51214\\
  0&0.53738&0.54099&0.54901&0.51487&0.51017&0.50279\\
  1&0.53554&0.53672&0.54693&0.49815&0.50623&0.49384\\
  2&0.53374&0.53211&0.54473&0.48183& 0.5024&0.48562\\
  3&0.53148&0.52714&0.54204&0.46604&0.49851&0.47834\\
  4&0.52845&0.52187&0.53859&0.45121&0.49433&0.47201\\
  5&0.52458& 0.5164&0.53424&0.43786&0.48969&0.46655\\
  6&0.51997&0.51082&0.52903&0.42632&0.48453&0.46176\\
  7& 0.5148&0.50527&0.52313&0.41665&0.47894& 0.4575\\
  8&0.50928&0.49982& 0.5168&0.40869&0.47309&0.45362\\
  9&0.50365&0.49458&0.51035&0.40218&0.46718&0.45002\\
 10&0.49808& 0.4896&0.50401&0.39684&0.46138&0.44664\\
 \hline\hline
\end{tabular}
\caption{The $h(q)$ values of Gaussian white noise (of data length
$1000$) calculated through wavelets from Daubechies family and
MF-DFA.}
\end{table}

\begin{table}
\centering
\begin{tabular}{|c|c|c|c|c|c|c|}
  \hline\hline
   q & Db4 & Db6 & Db8 & Linear & Quadratic & Cubic\\
  \hline\hline
-10&0.55498&0.52839&0.54909& 0.5331& 0.5637&0.59861\\
 -9&0.55122& 0.5249&0.54704&0.53022&0.55758&0.58947\\
 -8&0.54716&0.52141&0.54494& 0.5276&0.55133&0.57984\\
 -7&0.54282&0.51797& 0.5428&0.52539&0.54511&0.56988\\
 -6&0.53822&0.51464& 0.5406&0.52371&0.53909&0.55977\\
 -5&0.53343&0.51141&0.53832&0.52267&0.53343&0.54968\\
 -4&0.52853&0.50829&0.53594&0.52225&0.52825&0.53979\\
 -3& 0.5236&0.50524&0.53345&0.52226&0.52361& 0.5302\\
 -2&0.51874&0.50221&0.53083&0.52231& 0.5195&  0.521\\
 -1&  0.514&0.49913&0.52808&0.52183&0.51581&0.51223\\
  0&0.50944&0.49593&0.52518&0.52025& 0.5124&0.50387\\
  1&0.50506&0.49253&0.52214&0.51722&0.50907&0.49595\\
  2&0.50083&0.48886&0.51896&0.51271& 0.5056&0.48845\\
  3&0.49668&0.48485&0.51566&0.50707&0.50174&0.48141\\
  4&0.49253&0.48045&0.51225&0.50076&0.49728&0.47482\\
  5& 0.4883&0.47565&0.50875&0.49425&0.49208&0.46866\\
  6&0.48398&0.47046& 0.5052&0.48785&0.48611&0.46286\\
  7&0.47957&0.46495&0.50162&0.48172&0.47948&0.45739\\
  8&0.47512&0.45923&0.49805&0.47593&0.47243&0.45221\\
  9&0.47072&0.45345&0.49453& 0.4705&0.46525&0.44732\\
 10&0.46643&0.44776&0.49109&0.46543&0.45818&0.44274\\
  \hline\hline
\end{tabular}
\caption{The $h(q)$ values of Gaussian white noise (of data length
$5000$) calculated through Daubechies 4 (Db-4) to Daubechies 8
wavelets and MF-DFA.}
\end{table}

\begin{table}
\centering
\begin{tabular}{|c|c|c|c|c|c|c|}
  \hline\hline
  q & Db4 & Db6 & Db8 & Linear & Quadratic & Cubic\\
  \hline\hline
-10&0.62408&0.56361&0.56012&0.55754&0.58706&0.58575\\
 -9&0.61137&0.55897&0.55515&0.55431&0.57652&0.58211\\
 -8&0.59803&0.55425&0.55031& 0.5511&0.56615& 0.5787\\
 -7&0.58486&0.54953&0.54562&0.54805&0.55607&0.57562\\
 -6&0.57282&0.54489&0.54108& 0.5454&0.54631&0.57298\\
 -5&0.56253&0.54043&0.53669&0.54342&0.53685&0.57083\\
 -4&0.55405&0.53623&0.53244& 0.5425&0.52759&0.56919\\
 -3&0.54705&0.53236&0.52831&0.54308&0.51845&0.56802\\
 -2&0.54104&0.52891&0.52427&0.54554&0.50934& 0.5672\\
 -1&0.53565& 0.5259&0.52031&0.55009&0.50025&0.56657\\
  0&0.53055&0.52336&0.51644&0.55644&0.49128& 0.5659\\
  1&0.52554&0.52128&0.51263&0.56367&0.48267&0.56495\\
  2&0.52043&0.51963& 0.5089&0.57032&0.47483&0.56351\\
  3&0.51507&0.51835&0.50525&0.57505&0.46834&0.56138\\
  4&0.50933&0.51737& 0.5017&0.57717&0.46375&0.55845\\
  5&0.50314&0.51662&0.49824&0.57678&0.46144&0.55471\\
  6&0.49648&  0.516&0.49488&0.57446&0.46132&0.55023\\
  7&0.48944&0.51543&0.49164&0.57092&0.46277&0.54513\\
  8&0.48217&0.51482&0.48851&0.56675&0.46493&0.53964\\
  9&0.47485&0.51412&0.48552&0.56236&0.46699&0.53395\\
 10&0.46768&0.51331&0.48267&0.55802&0.46847&0.52826\\
  \hline\hline
\end{tabular}
\caption{ The $h(q)$ values of Gaussian white noise (of data length
$10000$) calculated through Daubechies 4 (Db-4) to Daubechies 8
wavelets and MF-DFA.}
\end{table}

\begin{table}
\centering
\begin{tabular}{|c|c|c|c|c|c|c|}
  \hline\hline
   q & Db4 & Db6 & Db8 & Linear & Quadratic & Cubic\\
  \hline\hline
-10&0.50973&0.50575&0.50463&0.59746&0.51373&0.49263\\
 -9&0.50728&0.50268&0.50228&0.58828&0.50577&0.49219\\
 -8&  0.505& 0.4998&0.50021&0.57966&0.49807&0.49214\\
 -7& 0.5029&0.49718&0.49839&0.57196&0.49083&0.49261\\
 -6&0.50099&0.49486& 0.4968&0.56548&0.48421&0.49371\\
 -5&0.49922&0.49291&0.49544&0.56046&0.47833&0.49556\\
 -4&0.49756&0.49138&0.49427&0.55703&0.47332&0.49827\\
 -3&0.49599&0.49032& 0.4933&0.55517&0.46924&0.50194\\
 -2&0.49451&0.48979&0.49251&0.55478&0.46613&0.50674\\
 -1&0.49315&0.48983&0.49191&0.55561&0.46399&0.51299\\
  0&0.49196&0.49047&0.49149&0.55732&0.46278&0.52127\\
  1&0.49098&0.49167&0.49126&0.55941& 0.4624&0.53257\\
  2&0.49027&0.49335&0.49119&0.56124&0.46269&0.54818\\
  3&0.48983&0.49536&0.49126&0.56217&0.46346&0.56935\\
  4&0.48967& 0.4975&0.49142&0.56172&0.46451&0.59634\\
  5&0.48975&0.49955& 0.4916&0.55975&0.46563& 0.6275\\
  6&0.48998&0.50133&0.49172&0.55647&0.46665&0.65973\\
  7&0.49028&0.50269& 0.4917&0.55225&0.46748&0.68996\\
  8&0.49057&0.50356&0.49147&0.54745&0.46805&0.71637\\
  9&0.49077&0.50396&0.49099& 0.5424&0.46836&0.73842\\
 10&0.49085&0.50391&0.49024&0.53731&0.46842&0.75633\\
\hline\hline
\end{tabular}
\caption{The $h(q)$ values of Gaussian white noise (of data length
$50000$) calculated through Daubechies 4 (Db-4) to Daubechies 8
wavelets and MF-DFA.}
\end{table}

\begin{table}
\centering
\begin{tabular}{|c|c|c|c|c|c|c|}
  \hline\hline
  q & $h(q)_{IC}$ & $h(q)_{IC_s}$ & $h(q)_{ISC}$ & $h(q)_{ISC_s}$ & $h(q)_{FP}$ & $h(q)_{FP_s}$\\
  \hline\hline
-10 &0.7233 &0.5325 &0.7308 &0.5416 &0.6684 &0.5309\\
-9  &0.7178 &0.5284 &0.7243 &0.5383 &0.6669 &0.5282\\
-8  &0.7119 &0.5244 &0.7173 &0.5350 &0.6655 &0.5254\\
-7  &0.7055 &0.5205 &0.7097 &0.5317 &0.6643 &0.5224\\
-6  &0.6987 &0.5169 &0.7017 &0.5285 &0.6632 &0.5193\\
-5  &0.6916 &0.5134 &0.6936 &0.5254 &0.6616 &0.5162\\
-4  &0.6844 &0.5102 &0.6857 &0.5224 &0.6585 &0.5130\\
-3  &0.6771 &0.5072 &0.6781 &0.5194 &0.6526 &0.5097\\
-2  &0.6698 &0.5044 &0.6710 &0.5166 &0.6424 &0.5065\\
-1  &0.6625 &0.5018 &0.6644 &0.5138 &0.6279 &0.5034\\
0   &0.6553 &0.4994 &0.6580 &0.5109 &0.6105 &0.5003\\
1   &0.6482 &0.4972 &0.6518 &0.5081 &0.5921 &0.4973\\
2   &0.6410 &0.4951 &0.6458 &0.5052 &0.5743 &0.4943\\
3   &0.6337 &0.4932 &0.6401 &0.5024 &0.5579 &0.4915\\
4   &0.6262 &0.4915 &0.6347 &0.4995 &0.5435 &0.4887\\
5   &0.6186 &0.4898 &0.6297 &0.4967 &0.5311 &0.4860\\
6   &0.6110 &0.4883 &0.6251 &0.4939 &0.5205 &0.4833\\
7   &0.6035 &0.4869 &0.6209 &0.4911 &0.5116 &0.4805\\
8   &0.5963 &0.4855 &0.6171 &0.4884 &0.5043 &0.4778\\
9   &0.5896 &0.4842 &0.6137 &0.4857 &0.4981 &0.4750\\
10  &0.5833 &0.4829  &0.610 &0.4831 &0.4929 &0.4723\\
  \hline\hline
\end{tabular}
\caption{ The $h(q)$ values of time series and shuffled time series
of i) ion saturation current (IS), ii) ion saturation current, when
the probe is in the limiter shadow (ISC) and iii) floating
potential, 6mm inside the main plasma (FP). Here the subscript 's'
refers values of shuffled time series. The Hurst measure is closer
to Brownian motion (H=0.5) for the shuffled times series, where as
for the unshuffled case, the presence of long range correlation
brings in substantial deviations.}
\end{table}
\section{Conclusion}
In conclusion, we have presented a reliable discrete wavelet based
method for estimating correlation and multiscaling behavior. This
approach is quite efficient and accurate in characterizing the
behavior of diverse non-stationary time series. Its efficacy is
derived from the optimal window size of discrete wavelet basis. A
single wavelet from Daubechies family is found to be good for
analyzing both the small and large fluctuations. The averaging over
the fluctuations from forward and backward procedure took good care
of both small and large fluctuations.

{\bf Acknowledgements} We would like to thank Dr. R. Jha for
providing the tokamak plasma data for analysis and Ms. Namrata Shah
for helping us in preparing this manuscript.

\end{document}